# Numerical calculations of radiative and non-radiative relaxation of molecules near metal particles

Maxim Sukharev[1] and Abraham Nitzan[2]

[1] Science and Mathematics Faculty, School of Letters and Sciences, Arizona State University, Mesa, Arizona 85212, USA
email: maxim.sukharev@asu.edu

[2] School of Chemistry, Tel Aviv University, Tel Aviv, 69978, Israel
email: nitzan@post.tau.ac.il

## Abstract

The dependence of the radiative emission and the non-radiative (energy transfer to the metal) relaxation rates of a molecule near a small metal particle on the molecule-to-particle distance and on the molecular orientation is calculated using a numerical solution of the Maxwell equations for a model that described the metal as a dispersive dielectric particle and the molecule as an oscillating point dipole. The emission rate is obtained by evaluating the total oscillating dipole in the system, while the non-radiative rate is inferred from the rate of heat production on the particle. For the distance dependence of the non-radiative rate we find, in agreement with experimental observations, marked deviation from the prediction of the standard theory of fluorescence resonance energy transfer (FRET). In departure from previous interpretations, we find that electromagnetic retardation is the main source of this deviation at large molecule-particle separations. The radiative emission rate reflects the total dipole induced in the molecule-particle system, and its behavior as function of distance and orientation stems mostly from the magnitude of the oscillating polarization on the metal particle (which, at resonance, is strongly affected by plasmon excitation), and from the way this polarization combines with the molecular dipole to form the total system dipole.

## 1. Introduction

Interest in the way by which proximity to dielectric interfaces affects the optical response of molecules goes back a long time. Such effects stem from the local properties of the electromagnetic field as well as the interaction of the molecular charge distribution with the dielectric environment. Renewed interest in such phenomena has followed the observation of surface enhanced Raman scattering nearly three decades ago, and the more recent emergence of molecular plasmonics as a field of fundamental and technological importance. The focus of the present paper is the phenomenon of energy transfer between an excited molecule and a nearby metal particle. This process has received considerable attention fueled by experimental observations of this phenomenon,(see[1,2] for early work and[3,4] for recent reviews) taken with a suitable theoretical analysis,[5-22] provide not only fundamental insight on the nature of molecular electronic relaxation near metal interfaces, but also a potential probe of the molecule-to-particle distance as well as the molecular orientation with respect to the particle surface;[12,23-30] for a review see Ref. 4. Using Forster resonance energy transfer (FRET) as a distance ruler in molecular systems has indeed been a very useful technique with many applications in studies of chemical and biological processes.[31-34] Such applications are based on the well-characterized and well-understood $R^{-6}$ behavior of this phenomenon, where $R$ is the distance between the donor and acceptor of excitation energy.

Applications of such energy transfer techniques to molecule-surface and molecule-particle systems such as those mentioned above have led to some confusion. Focusing on a system comprising of a single molecule in proximity to a single metal nano-particle the process of energy transfer from the molecule to the particle is theoretically understood and has been treated by several classical, semiclassical and quantum approaches.[5-11,13-22] Interpretations of experimental observations are sometimes at odds with these calculations. Some of the observed disagreements can be attributed to the oversimplified details of the theoretical models. For example, the latter usually employ a single point dipole to represent the molecular system while many experiments involve large molecules or a molecular layer. Also, the non-local character of the metal dielectric response may affect the energy transfer for molecule-to-particle distance smaller than 2-3 nm. Although theoretical treatments of such non-local effects have been advanced, [35,11,36] most practical calculations rely on the much simpler model of local dielectric response. An issue of arguably more significant importance is the dependence of the energy transfer efficiency on the molecule-to-particle distance that pertains directly to the important application of this phenomenon as a distance ruler as used in molecular FRET. In contrast to the $R^{-6}$ dependence that characterizes the FRET phenomenon, a weaker $R^{-n}$ with $n \sim 4$ is often reported,[3,4,27,28,37-41] although deviations from both behaviors are also observed.[29]

These observations were interpreted[3,4,27,28,37-41] as evidence for the dominance "surface energy transfer" (SET) mechanism following the theoretical prediction[10] that the transfer of molecular electronic excitation to flat metal surfaces should satisfy an $R^{-4}$ dependence on the molecule-to-surface distance. Such an interpretation, however, contradicts our basic understanding of the origin of these distance scaling behaviors. Evidently, the $R^{-6}$ dependence of FRET between molecular species originates form the $R^{-3}$ dependence of the dipole-dipole interaction, which is usually the dominant coupling for such process, with the corresponding rate scaling as the coupling squared. When the energy is transferred to the volume of a bulk solid, the rate should be summed over all accepting species, that is, integrated over the solid volume; hence the distance scaling changes from $R^{-6}$ to $R^{-3}$. However, as explained in Ref. 10 (see also Ref. 12), near a flat surface momentum conservation restricts energy transfer to the solid surface only, whereupon integration over surface yields a $R^{-4}$ dependence. All this, however, is relevant for a molecule situated near a flat surface or, for as a reasonable approximation for a molecule-particle system, when the molecule-to-particle distance is far smaller than the particle radius, and is indeed observed when these conditions are met (see, e.g. Ref. 30 for a recent example). The experiments referred to above, however, involve molecule-to-particle distances comparable or larger, sometimes much larger, than the particle sizes. Their observed distance dependence cannot be classified as a SET process as described in Ref. 10.

This discrepancy between experimental observations and theoretical understanding has already been pointed out.[14-17] and the observed $R^{-3} - R^{-4}$ dependence was rationalized either as and intermediate distance behavior[14-17,42] or as an indication that the donor-acceptor distance has not been inferred correctly from the molecular structure.[26] It should be emphasized that the calculations in Refs. 14-17,42 where done in the electrostatic limit.

In the present paper we examine this issue without the limitation of the electrostatic limit, by a direct numerical integration of the Maxwell equations in a system comprised of a molecule, represented by a classical point dipole, and a metal particle characterized by a given (local) dispersive dielectric response, using the computed electric field and polarization distributions to evaluate radiative and non-radiative[43] relaxation rates as described in Refs. 7, 44. This yields the dependence of these rates on the molecule-to-particle distance, particle size (and potentially shape, however in this study only spherical particles are considered) and molecular orientation. With respect to the dependence of the molecule-to-metal energy transfer rate on the molecule-to-particle distance, our results support the experimental observations described above, however our main conclusion is that for distances larger than the particle size this dependence not associated with a SET process as analyzed in Ref. [10]. Still, for particle of radius larger than ~ 1 nm the distance dependence $R^{-6}$ typical to FRET is not observed. At close proximity (relative to the particle size) the particle appears as a nearly flat surface, yielding an $R^{-n}$; $n = 3...4$ dependence, [NOTE: The qualitative argument is that at very close proximity the molecule sees a flat surface, therefore $n = 4$, while at larger distance where the surface curvature has an effect, $n = 3$ is more appropriate, however the distance range where such arguments are valid is too small for the distinction between these behavior to become apparent.] while at larger distances retardation effects cannot be disregarded. In the asymptotically large distance limit (the far field or the radiation zone) such effects result in $R^{-2}$ distance dependence, as expected in the far field (radiation) regime. For particle sizes larger than ~ 10 nm we find that rather than observing a distance dependence $R^{-n}$ with $n$ changing from 3…4 to 6 and finally to 2, we find that the $R^{-6}$ behavior is not realized at all. Instead, the distance dependence depends on the molecular orientation relative to the particle surface, and can be easily fit to an extended $R^{-4}$ regime as observed experimentally and as can be rationalized from the analytical expression for the field of a radiating dipole.

At the same time we also examine the radiative (fluorescence) rate. We find that fluorescence from the system can be either enhanced or damped relative to the emission rate by an isolated excited molecule, depending on the molecule orientation relative to the particle surface and in agreement with experimental observations.[45] The fluorescence quantum yield is reduced at close proximity to the surface but can be considerable enhanced at larger distances.

The behavior of excited molecules (modeled as classical oscillating dipoles) near metal nanostructures was recently explored by other research groups. These studies mostly focused on specific geometries of the metal sub-system such as nanoantennas,[46] metal-dielectric-metal waveguide structures,[47] or nanoparticle dimers.[48] Here we employ rigorous numerical simulations to re-examine the optical response of a simpler structure, a molecule near a spherical metal nanoparticle, in order to clarify the issues described above.

Section 2 describes our numerical procedure. In Section 3 we present our results for the behavior of the radiative and non-radiative relaxation of an excited molecule near a metal particle, focusing on the molecule particle distance as described above, but also on molecular orientation and particle size. Section 4 concludes.

## 2. The computational model

To simulate the optical response of a molecule coupled to a metallic particle we employ three-dimensional fully vectorial electromagnetic model using classical Maxwell's equations

$$\frac{1}{c}\frac{\partial \vec{E}}{\partial t} = \nabla \times \vec{H} - \vec{J}, \qquad (1a)$$

$$\frac{1}{c}\frac{\partial \vec{H}}{\partial t} = -\nabla \times \vec{E}, \qquad (1b)$$

where $c$ is the speed of light in vacuum. The density current, $\vec{J}$, in Eq. (1a) represents the currents in spatial regions occupied by the metal (see below) as well as the oscillating current associated with a point dipole that represents the molecule, which drives the system. To take into account the dielectric dispersion of a metal we use the Drude model with the dielectric function written in the frequency domain as

$$\varepsilon(\omega) = \varepsilon_R - \frac{\Omega_p^2}{\omega^2 - i\Gamma\omega}, \qquad (2)$$

here $\varepsilon_R$ is the asymptotic value of the dielectric constant at high frequencies, $\Omega_p$ is the bulk plasma frequency, and $\Gamma$ is the phenomenological damping. In the time domain the Ampere law (1a) with dispersion in the form of (2) results in an additional equation that defines the dynamics of the density current [49]

$$\frac{\partial \vec{J}}{\partial t} + \Gamma \vec{J} = \frac{\Omega_p^2}{4\pi} \vec{E}. \tag{3}$$

In the calculations described below the metal particle is taken spherical and, unless otherwise stated, the sphere radius is 20 nm. The following set of parameters was used for silver: $\varepsilon_R = 8.926$, $\Omega_p = 1.760 \times 10^{16}$ rad/sec, $\Gamma = 3.084 \times 10^{14}$ 1/sec. Simulations for gold particles were performed with parameters: $\varepsilon_R = 9.500$, $\Omega_p = 1.360 \times 10^{16}$ rad/sec, $\Gamma = 1.048 \times 10^{14}$ 1/sec. The dielectric constant outside the metal is taken 1.

The calculations are performed in the steady state mode.[7,44] In this approach the excited molecule, represented by a point dipole $\vec{\mu}(t)$ [50] that oscillates with constant amplitude $\vec{\mu}$ at the molecular transition frequency $\omega$, and the steady state response of the system is evaluated. The response function relevant to the present study is the amplitude $\vec{E}(\mathbf{r})$ of the local electric field at position **r** in the system that oscillates at the driving frequency. As discussed below, both the radiative relaxation rate and the molecule-to-particle energy transfer rate can be evaluated from this field. Note that the molecular excitation process is disregarded in this calculation since in the linear regime considered it does not affect the calculated rates. It should be kept in mind however that optical response properties of molecules near metal nanostructures may stem from other processes such as those originating from the effect of the exciting field on the metal structure, in particular at frequencies close to surface plasmon-polariton resonances in these structures.

We integrate the corresponding system of equations (1) and (3) with a pointwise driving dipole source placed at some distance with specific orientation relative to the metal sphere. The numerical integration is performed utilizing finite-difference time-domain method (FDTD)[51] using home-built codes. Space is discretized employing Yee's algorithm and electromagnetic fields are propagated in time via a leapfrog time-stepping technique. Open boundaries are simulated using convolution perfectly matched layers (CPML) absorbing boundaries.[52] A standard way of calculating the radiative and non-radiative relaxation rates of a molecule[53] is to calculate total power radiated by the molecular dipole and the power radiated by this dipole to the far field. The latter is usually simulated by evaluating an integral

$$P_{rad} = \oint \vec{S}\vec{n}\,ds, \tag{4}$$

over a closed surface encompassing the system, where $\vec{S}$ is the Poynting vector and $\vec{n}$ is the normal to the surface. In principle integral (4) has to be evaluated in the far field at a distance much larger than characteristic size of the system. Numerically however this could be simulated using near-field-to-far-field transformation technique.[51] Even in the latter case one has to carefully place the integrating surface far enough such that any possible evanescent fields supported by the system are not contributing to the integral (4).

Here we employ more efficient and less numerically expensive method. It is possible to show[7,44] that the radiative decay rate is determined by the total dipole moment of the system molecule + particle, $\vec{\mu}_{total} = \vec{\mu}_{molecule} + \vec{\mu}_{particle}$, and can be calculated in the far-field according to

$$\Gamma_r = \frac{\omega^3}{3\hbar c^3}\left|\vec{\mu}_{total}\right|^2. \tag{5}$$

The induced dipole moment of the particle is calculated using

$$\vec{\mu}_{particle} = \frac{\varepsilon - 1}{4\pi}\int \vec{E}\,dV, \tag{6}$$

where the integral is taken over particle's volume.

The non-radiative decay rate can also be calculated from the local electric field in the particle. Indeed, the dissipation rate in the particle is given by

$$\Gamma_{nr} = \frac{\mathrm{Im}(\varepsilon)}{8\pi\hbar}\int \left|\vec{E}\right|^2 dV. \tag{7}$$

with the integral taken again over particle's volume. The emission yield is then given by

$$Y_r = \frac{\Gamma_r}{\Gamma_r + \Gamma_{nr}} \tag{8}$$

It should be noted that $\Gamma_{nr}$, Eq. (7), is not the same as the rate of energy transfer rate from the molecule to the metal particle because the excited metal particle may still decay radiatively, however at any frequency the two rates are proportional to each other, therefore $\Gamma_{nr}$ can be used to evaluate the distance dependence of the energy transfer rate as done in the next section. [NOTE: Note that because we cannot usually distinguish between emission from the particle and from the molecule, Eq. (8) is the relevant expression for the quantum yield]

In the linear regime considered, the radiative and non-radiative rates can be evaluated also using the short-pulse method (SPM). In this method, the molecular dipole is represented as an ultra-short pulse with a bandwidth essentially flat in the frequency range of interest (we used 0.36 fs pulses in SPM simulations) and the resulting frequency dependent response is obtained by Fourier transform, making it possible to obtained response at many frequencies from a single FDTD run. This is particularly simple for the radiative rate since the induced dipole (6) is proportional to electric field. We excite the system molecule + particle with a short molecular dipole pulse. Maxwell's equations are then propagated for approximately 1 ps to insure numerical convergence, and the integral (6) is evaluated as function of time during this evolution. Once the propagation is complete we perform fast Fourier transform of the stored data.

Due to nonlinear dependence of $\Gamma_{nr}$ on the electric field amplitude, application of the SPM is more complicated in this case because the calculations of (7) have to be carried out independently at different frequencies. SPM could still be used if the field could be stored at each spatial position, but memory limitation prohibits such a procedure. Instead, it is possible to perform the necessary Fourier transform of the volume integral $\int |\vec{E}|^2 dV$ on the fly within a single FDTD run. This is accomplished via calculations of so-called phasor functions,[51] which essentially are related to steady-state solutions of corresponding Maxwell's equations. Although only single FDTD run is necessary for these calculations it is still noticeably slower compared to the simple SPM method.

The numerical convergence of all results presented in the next Section is achieved with the spatial resolution of 2 nm. We also note that simulations are carried out using high value of a molecular transition dipole of $10^3$ C m. This is done in order to ensure that computed EM fields are higher than known cut-off values for FORTRAN. Since our calculations are performed in the linear regime, this choice has no physical consequence. Also, the results presented in the next section are normalized with respect to the radiative decay rate of the free molecule and as such are not affected by this choice.

The next Section discusses the results of our simulations where the main focus is on the dependence of radiative and non-radiative decays rates (5) and (7) on the distance between the metal particle and the molecule.

## 3. Results and discussion

Fig. 1 displays the basic optical properties of silver and gold spheres (radius 20 nm), in terms of their optical response to an incident plane electromagnetic wave. Fig. 1 shows both the normalized scattering intensity and the normalized total absorption cross-section as functions of the incident frequency. The former is evaluated by calculating the Poynting vector at a given detection point as indicated in the inset of Fig. 1a. The later is obtained under the assumption that the absorption cross-section is proportional to the heat developed in the particle (as obtained from Eq. (7)). Both figures show the prominent peaks near 3.5 eV (for silver) and 2.6 eV (for gold) that characterizes the dipolar plasmon excitation in a sphere of the corresponding metal. The very slight shifts seen between the scattering and absorption peaks reflect the different frequency dependencies associated with Eqs. (6) and (7).

The other figures pertain to the behavior of an excited molecule situated near a metal particle. We have studied two configurations with the molecular dipole perpendicular or parallel to the particle surface (see the inset in Fig. 2). Figures 2 and 3 show the non-radiative relaxation rates, normalized by the radiative rate of the free molecule, as functions of the molecule-to-particle's surface distance, for the case where the frequency is far from (Fig. 2) and near (Fig. 3) the plasmon resonance. Fig. 4 compares the non-radiative relaxation rates of a molecular dipole in the parallel configuration near silver and gold spheres of radius 8 nm, calculated at the corresponding plasmon resonance frequencies. Finally, the radiative relaxation rate for the same geometries and frequencies is shown in figures 5 and 6. The insets to the latter figures show the corresponding quantum yields, Eq. (8).

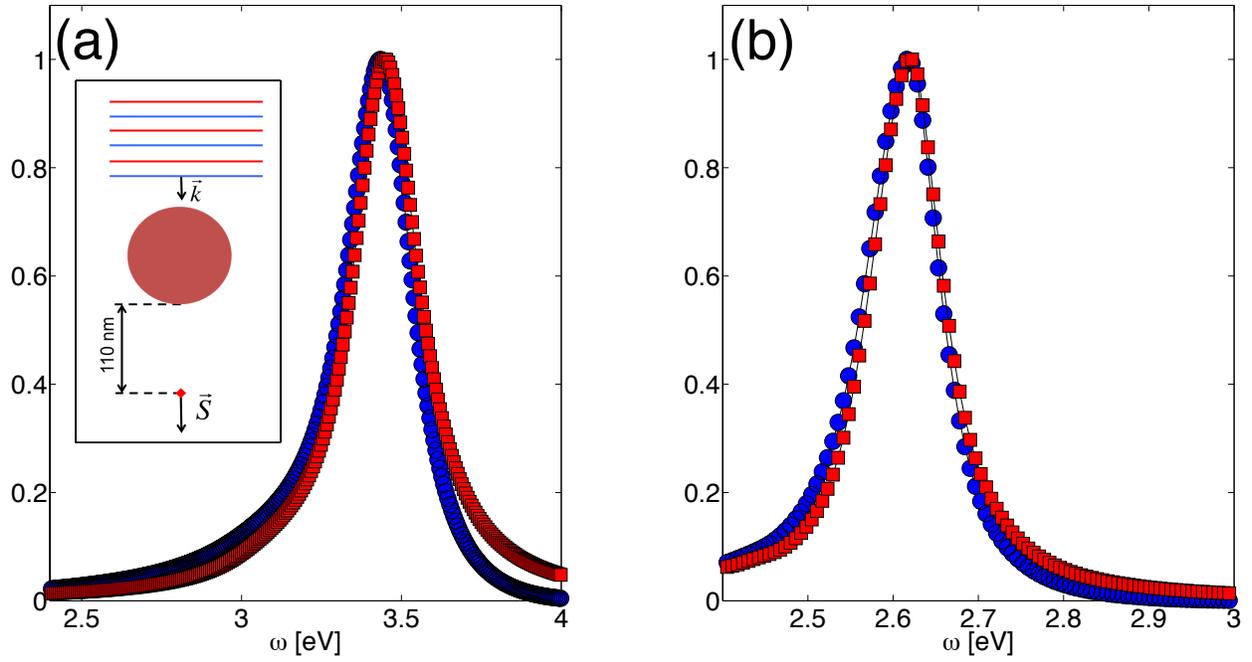

Fig. 1. (Color online) The normalized light scattering intensity (black circles) and the normalized total absorption cross-section (red squares) as functions of the incident frequency (in eV) evaluated for silver (panel (a)) and gold (panel (b)) spheres of radius 20 nm. The inset in panel (a) schematically shows calculations of the scattering intensity. Here the Poynting vector component parallel to the incident wave vector *k* is calculated using the scattered EM field at a given detection point indicated as a red dot in the inset.

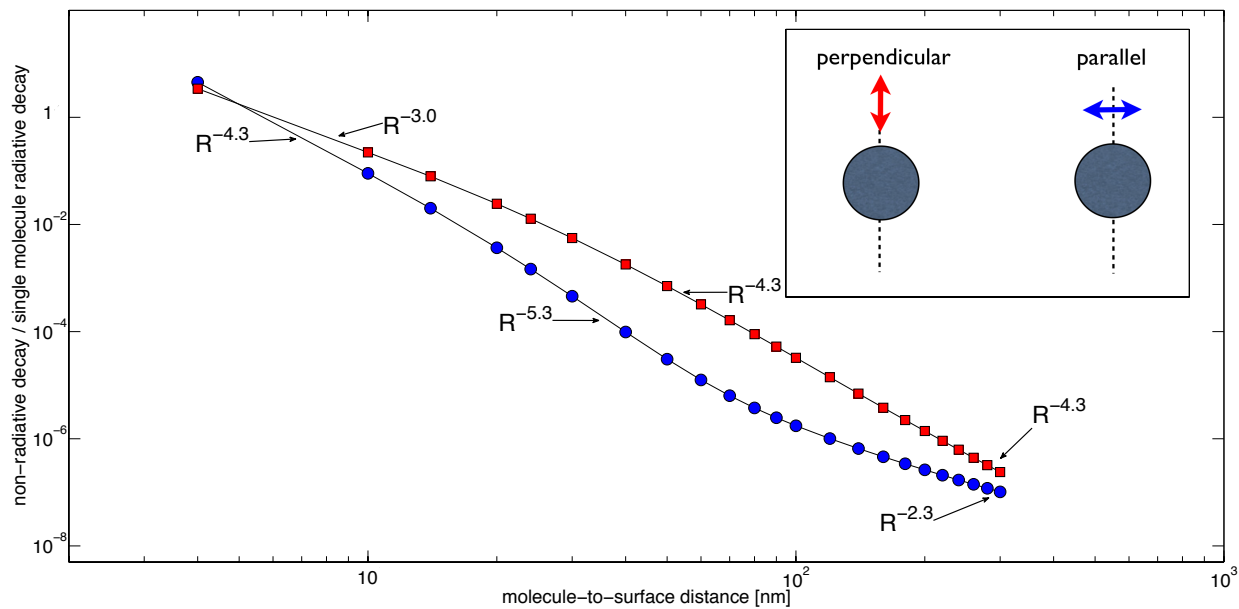

Fig. 2. (Color online) The non-radiative decay rates of a molecule characterized by transition frequency 1.0 eV, normalized by a radiative decay of a single molecule, shown as functions of the molecule-to-surface distance, *R*. Calculations were performed for silver nanoparticle of radius of 20 nm at two molecular orientations depicted in the inset. Blue circles and red squares show results for parallel and perpendicular orientations, respectively, of the molecular dipole relative to the sphere surface. The distance dependence of the rates is indicated by the local slopes shown for small, intermediate and large distances.

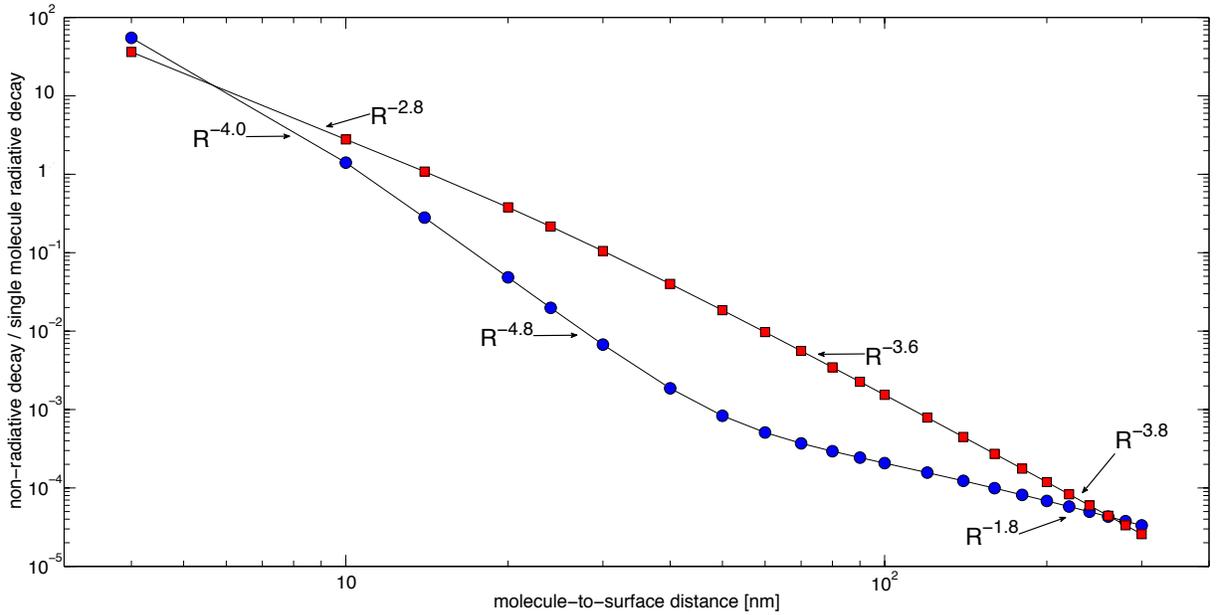

Fig. 3. (Color online) Same as Fig. 2, for a molecule characterized by transition frequency 3.41 eV - at the plasmon resonance of the silver nanoparticle.

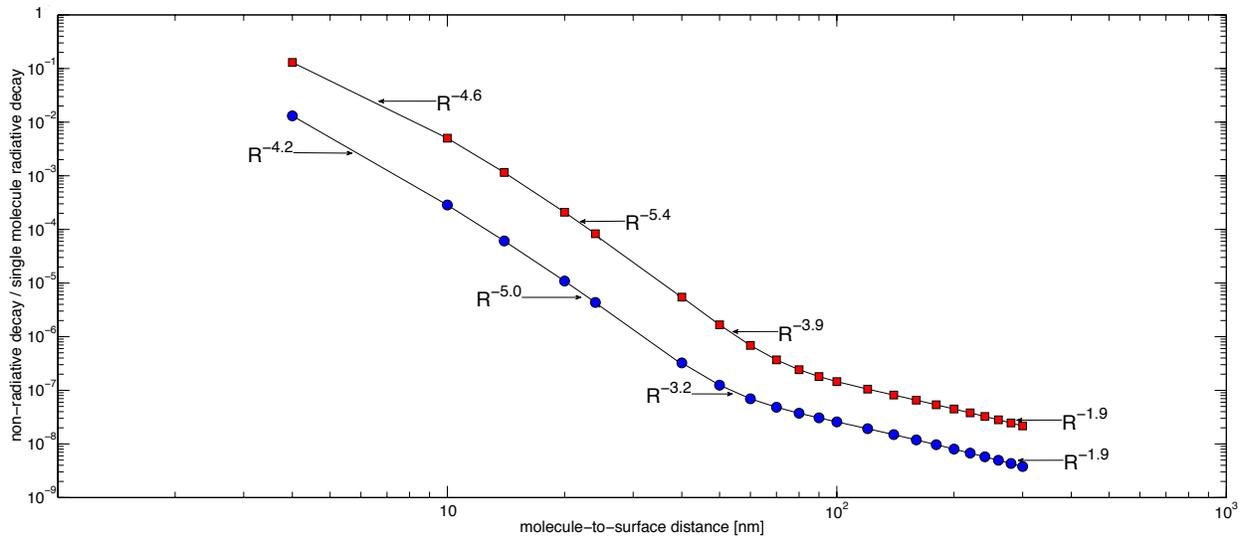

Fig. 4. (Color online) The non-radiative decay rates normalized to a single molecule decay rate as functions of molecule-to-surface distance, $R$. The two curves correspond to silver (blue circles) and gold (red squares) spheres of radius of 8 nm, with the molecule in parallel orientation (see inset in Fig. 2). The frequency of the molecular dipole oscillations that drive the system is taken at the corresponding plasmon resonances of each particle (3.41 eV for silver and 2.63 eV for gold).

Consider first the calculated non-radiative relaxation rates shown in Figs. 2-3. Several features are evident: (a) the distance dependence of the non-radiative rate is sensitive to the molecular orientation relative to the sphere surface; (b) at small (relative to the particle size) metal-to-particle separations, the distance dependence approaches $R^{-n}$; $n \sim 3-4$ that is characteristic of energy transfer to the bulk or surface of a macroscopic body; (c) the result for the parallel configuration (that corresponds to an effect of the dipole field in a direction normal to the dipole (see Eqs. (9)-(10)) approaches the $R^{-2}$ dependence, characteristic of the radiation regime, already at distance of order 100 nm; (d) the $R^{-6}$ dependence expected for FRET between molecules of sizes smaller than intermolecular distance is not realized here even at distances large relative to the sphere size. An approach to this behavior is seen at intermediate distances and it is more pronounced for smaller particles (Fig. 4).

As pointed out in Section 2, these non-radiative rates are proportional to the energy transfer rates that are usually addressed in the standard treatments of molecular energy transfer, so the distance dependence shown in Figs. 2 and 3 reflects that of the latter. Indeed, the computed distance dependence agrees with experimental observations.[3,4,27,28,37-41] Viewed within the standard theoretical treatments of FRET, this distance dependence appears surprising since such treatments predict an asymptotic $R^{-6}$ behavior. As outlined in Section 1, attempts to understand these observations have invoked surface energy transfer (SET), which is indeed characterized by an $R^{-4}$ distance dependence,[10] however only when the distances involved are much greater than the particle/molecule sizes. The calculations displayed in Figs. 2 and 3 show that the origin of the $R^{-4}$ dependence is different. It reflects the fact that for distances large relative to the particle size, the electrostatic (long wavelength) approximation used in the standard Forster theory does not hold and that retardation effects in the Maxwell equations need to be taken into account. This can be seen from considering the electric field at position **r** associated with an oscillating point dipole source[54] positioned at the origin

$$\mathbf{E}(\mathbf{r}) = k^2 (\mathbf{n} \times \vec{\mu}) \times \mathbf{n} \frac{e^{ikr}}{r} + \left[3\mathbf{n}(\mathbf{n} \cdot \vec{\mu}) - \vec{\mu}\right]\left(\frac{1}{r^2} - \frac{ik}{r}\right)\frac{e^{ikr}}{r} \qquad (9)$$

where **n** is a unit vector in the direction of **r** and $c$ is the speed of light in vacuum. For a dipole perpendicular and parallel to **r** this leads to

$$|E_\perp|^2 = \frac{k^4 |\vec{\mu}|^2}{r^2}\left[1 + \frac{3}{(kr)^2} + \frac{1}{(kr)^4} + ...\right] \quad (10a)$$

$$|E_\parallel|^2 = \frac{4k^4 |\vec{\mu}|^2}{r^2}\left(\frac{1}{(kr)^2} + \frac{1}{(kr)^4} + ...\right) \quad (10b)$$

Note that $E_\perp$ and $E_P$ corresponds to the parallel and perpendicular cases, respectively, in Figs. 2 and 3. While the field expressions (10) are that of a free dipole and the field in Eq. (7) corresponds to the dipole-sphere system, it is easily seen that the distance dependence in Eq. (10) is qualitatively manifested in Figures 2 and 3: in the electrostatic limit, $k \to 0$, both Eqs. (10) show a $r^{-6}$ dependence. However, for finite $k = \omega/c$ Eq. (10b) shows an asymptotic $r^{-4}$ behavior while Eq. (10a) shows an intermediate $r^{-4}$ dependence and an asymptotic $r^{-2}$ behavior. We conclude that the numerical results of Figs 2 and 3 as well as the experimental observations of Refs. 3,4,27,28,37-41 reflect these retardation effects in the Maxwell equations.

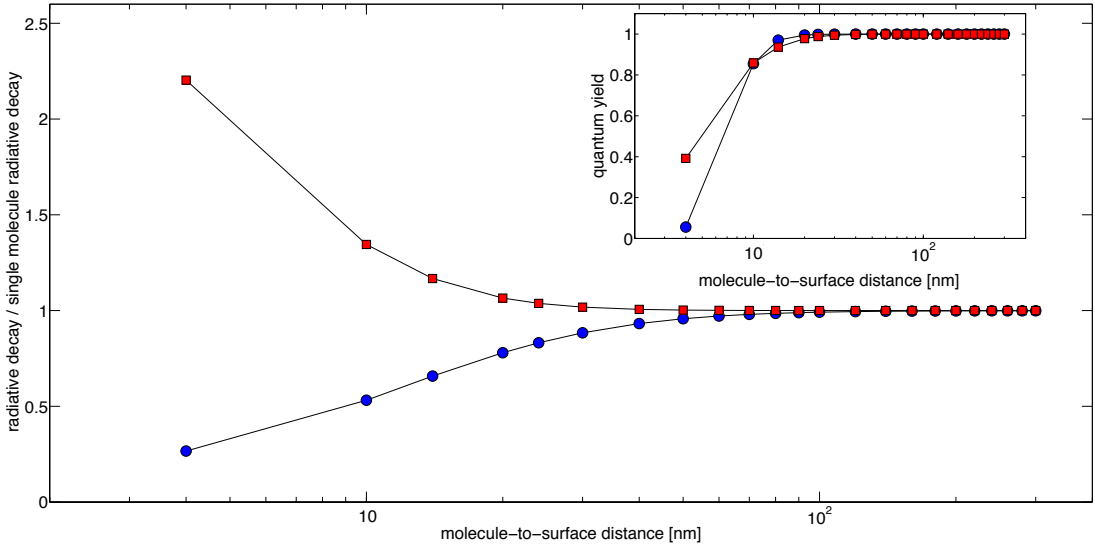

Fig. 5. (Color online) Radiative decay rates of a molecule characterized by transition frequency 1.0 eV, normalized by a radiative decay of a single molecule, shown as functions of the molecule-to-surface distance, $R$, (in nm). Calculations were performed for silver nanoparticle

with a radius of 20 nm at two molecular orientations depicted in the inset of Fig. 2. Blue circles and red squares show results for parallel and perpendicular orientations, respectively, of the molecular dipole relative to the sphere surface. The inset shows corresponding quantum yields calculated from Eq. (8).

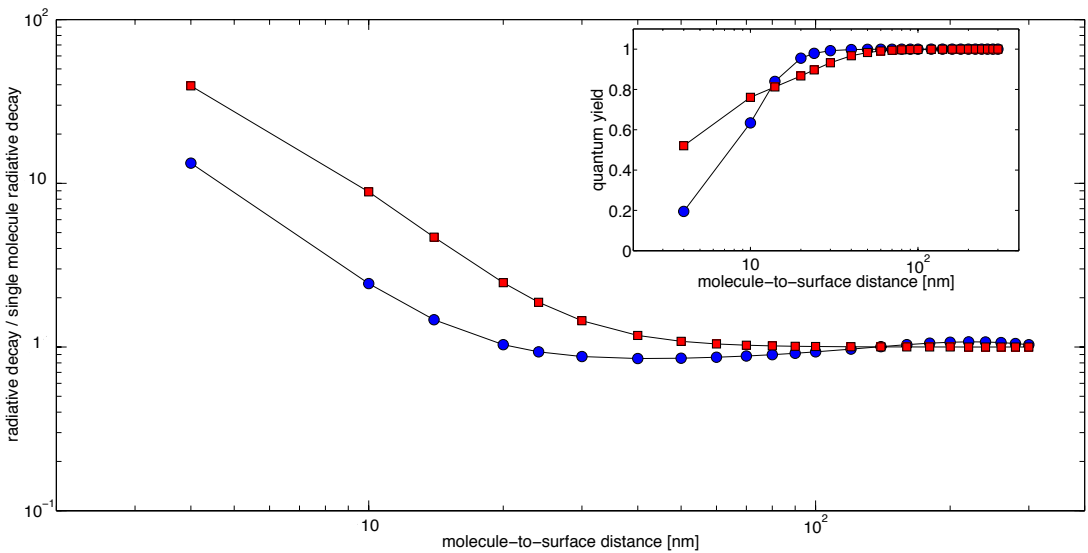

Fig. 6 (Color online) Same as Fig. 5, for a molecule characterized by the transition frequency 3.41 eV - at the plasmon resonance of the silver nanoparticle.

Next consider the radiative rates. Experimentally both enhanced fluorescence and fluorescence reduction have been reported. They are realized when the molecular dipole and the dipole induced on the metal particle combine constructively and destructively, respectively. This is clearly seen in Figs. 5 and 6. In this low incident frequency regime the metal response can be qualitatively estimated from the image induced on a perfect metal. At close proximity, where retardation effects are negligible, these dipoles add constructively in the perpendicular configuration and destructively in the parallel case. This is seen explicitly in the enhanced or reduced fluorescence rates displayed in Fig. 5. The situation is different when the molecular frequency is close to the plasmon resonance of the particle (Fig. 6). Fluorescence enhancement, resulting from the excitation of the particle plasmon, is seen in both geometries; still it is considerably larger in the perpendicular configuration. It is interesting to note that in this resonance case, the emission rate in the parallel configuration (blue line of Fig. 6) goes through a slight minimum (reflecting destructive interference between the dipole and its image) as function of distance, at a molecule-surface distance ~50 nm, before it increases at smaller separation. A much more pronounced minimum is seen with other choices of parameters[44] and was observed also experimentally.[23]

It is important to note, that although fluorescence can be enhanced for suitable configurations at close proximity to a particle surface, the emission yield always decreases at smaller molecule-particle distance (see insets to Figs. 5 and 6), indicating that the enhancement is stronger for the non-radiative relaxation due to energy transfer to the metal particle. Only at large enough distances the emission yield is close to unity, indicating that the radiative emission is larger than the non-radiative relaxation. At such distances however enhancement relative to the isolated molecule is small. It is of interest to explore possible configurations (necessarily at intermediate distances) where emission is enhanced without being compromised too much by the metal induced non-radiative decay of the emitter.

## 4. Conclusions

We have studied the distance dependence of radiative emission and non-radiative (energy transfer to metal) rates of an excited molecule as a function of distance from a small metal particle. As described in Section 2, both effects are related to the behavior of the local electromagnetic field in the space occupied by the metal particle and reflect plasmon excitions and dielectric damping in this space. In addition the emission rate is affected by the way (constructive or destructive) by which the molecular dipole and the dipole induced on the particle combine to form the total system dipole.

Regarding the non-radiative relaxation, experimental results that indicate qualitative deviation from the FRET theory of energy transfer where corroborated by the numerical calculation, however for large molecule particle distances the origin of these deviations is found to result from retardation corrections to the long-wavelength limit considered in the FRET theory and not by a surface energy transfer mechanism as suggested by earlier works. Surface energy transfer dominates at distances much smaller than the particle size while at large distances the characteristic $R^{-2}$ behavior of the far field is approached. Interestingly, for the particles sizes (8 and 20 nm) considered in our calculation the FRET distance dependence ($R^{-6}$) is not observed even at intermediate distances because the needed condition, $a = R = \lambda$ (where $a$ is the particle size and $\lambda$ is the radiation wavelength) is not sufficiently realized.

The emission rate reflects the total dipole on the molecule-particle system. As already stated, its dependence on the metal-particle separation and orientation stems from the way (constructive or destructive) by which the molecule and particle dipoles combine. Again, for large distances, retardation effects should be taken into account in a full calculation of this effect.

The model used in this calculation is idealized in several respects. First, the local bulk dielectric function is used to describe the optical response of the metal particle. Second, the molecule is represented by a classical point dipole. Third, a single molecule is considered. It is the latter idealization which, we believe, deserved the closest attention. A group of optically active molecules interaction with a plasmonic particles is a prototype model for a host of phenomena associated with plasmon-exciton interaction. The optical response of such systems is now an active field of study, but little is known about this response in the time domain, where, in addition to radiative and non-radiative relaxation, coherent response and decoherence rates should be taken into account. We plan to address such systems in future studies.

**Acknowledgements**. The research of A.N. is supported by the Israel Science Foundation, the Israel-US Binational Science Foundation and the European Science Council (FP7 /ERC grant no. 226628). M.S. is grateful to the Air Force Office of Scientific Research for the Summer Faculty Research Fellowship 2013.